\documentclass[conference,letterpaper]{IEEEtran}

\usepackage[utf8]{inputenc}
\usepackage[T1]{fontenc}
\usepackage{url}
\usepackage{ifthen}
\usepackage{cite}
\usepackage[cmex10]{amsmath}
\usepackage{amsthm}
\usepackage{amsfonts}
\usepackage{graphicx}
\usepackage{subcaption}
\newtheorem{theorem}{Theorem}

\newtheorem{defn}{Definition}

\begin{document}
\pdfoutput=1
\title{An Environmentally-Adaptive Hawkes Process with An Application to COVID-19 }


\author{%
  \IEEEauthorblockN{Tingnan Gong, Yu Chen and Weiping Zhang}
  \IEEEauthorblockA{Department of Statistics and Finance \\
                   University of Science and Technology of China\\
                    Hefei, Anhui, P.R. China\\
                    Email: cyu@ustc.edu.cn}
}


\maketitle

\begin{abstract}
   We proposed a new generalized model based on the classical Hawkes process with environmental multipliers,  which is called an environmentally-adaptive Hawkes (EAH) model.  Compared to  the classical self-exciting Hawkes process,  the EAH model exhibits more flexibility in a macro environmentally temporal sense,   and  can model  more complex processes by using dynamic branching matrix.  We  demonstrate the  well-definedness of this EAH model.  A more specified version of this new model is applied  to  model COVID-19 pandemic data through an efficient EM-like algorithm. Consequently, the proposed model consistently outperforms the classical Hawkes process.
\end{abstract}


\section{Introduction}
The self-exciting point process has been proposed in various fields such as epidemic forcasting (see \cite{meyer2012space}, \cite{meyer2014power}).  The Hawkes process is a classical self-exciting point process,  in which, each event $i$ occurs at time $t_i$, and generates new events with stochastic intensity $\phi(t-t_i)$, where $\phi(t)$ is a generic kernel. Thus Hawkes processes prove useful in modeling some real-world complex systems with clusters (see \cite{bacry2015hawkes}, \cite{hawkes2018hawkes}).   The branching matrix defined in a multivariate Hawkes process is responsible to describe the complex excitements between dimensions (nodes) in a network. Particularly, the  $(i,j)$ entry in the branching matrix is non-negative, measuring the excitement from node $j$ to node $i$. Whereas the branching matrix is static in a classical Hawkes process, the rigidity arises in a temporal sense.

Many of current work utilized the Hawkes process to fit the COVID-19 pandemic (see \cite{chiang2020hawkes}, \cite{escobar2020hawkes}. \cite{lesage2020hawkes}). With the self-exciting property, the Hawkes process is a seemingly proper tool to fit the COVID-19 spread. However, as shown in Section \ref{numerical exp}, the self-exciting property tends to build an explosive Hawkes process and overestimate the deterioration. In many areas of the real world, indoor public spaces were closed and people were required to wear a mask. Even many cities were locked down. Soon the spread of virus would slow down and exhibit not like a typical Hawkes process anymore. In this paper, we expect to apply our proposed model to the  COVID-19 pandemic to capture the time-varying epidemic control, which is brought by  the  alertness  of the society and the macro countermove of the local governments. Thus decaying environmental multipliers are considered when constructing the new model.

We develop an environmentally-adaptive Hawkes (EAH) model with dynamic multipliers. The dynamic multiplier is called the environmental multiplier, which is a time-varying function. By choosing proper environmental multipliers, the model exhibits various temporal flexibility.  The proposed EAH model is a novel and sophisticated version of the Hawkes model. We firstly provides a cluster representation of EAH model  to justify its well-definedness (existence). From the analysis of the existence, the sufficient condition ensuring the non-explosiveness of an EAH model is given. Then we state the uniqueness of the EAH model. Also based on \cite{li2017detecting}, an EM-like algorithm is developed to efficiently estimates the parameters in the environmentally-adaptive Hawkes model with decaying multipliers (EAHDM), a pandemic-specified version of the EAH model. 

The rest of this paper is organized as follows.  Section \ref{hawkes process} introduces the background knowledge on the Hawkes process. In Section \ref{EAH section}, we formally construct the EAH model and clarify its existence and uniqueness. Several properties related to the probability generation functional (p.g.fl. see \cite{vere1970stochastic}) are given. Also, the multi-dimensional EAH and EAHDM model are briefly introduced. In Section \ref{EM algo}, we develop the EM-like algorithm. In Section \ref{numerical exp}, we demonstrate the proposed methodology via both simulations and the COVID-19 data.

\section{Background}\label{hawkes process}
An $M$-dimensional Hawkes process is denoted by $\boldsymbol{H}(t)=(H^1(t), \cdots, H^M(t))^\top$, which can be fully characterized by its intensity $\boldsymbol{\eta} = [\eta_1(t), \eta_2(t), \ldots, \eta_M(t)]^\top$ as follows,
\begin{equation}\label{intensity of HP}
	\boldsymbol{\eta}(t) = \boldsymbol{\mu} + \int_0^t \left(\mathbf{A} \odot \Phi(t-\tau)\right)\mathrm{d}\boldsymbol{H}(\tau),
\end{equation}
where
$\odot$ represents the Hadamard (elementwise) multiplication  and $\boldsymbol{\mu}$ is the intensity vector of the immigrant process. The concept of the immigrant process (see  \cite{hawkes1974cluster}) will be further introduced in Section \ref{EAH section}.    $\mathbf{A}=(\alpha_{ij})_{M\times M}$ is the branching matrix. $[\Phi_{i,j}(t)]_{i,j = 1}^M$ is the kernel function matrix, or fertility function matrix, which provides a useful tool to characterize the influence of the history information and the effect from node $j$ to node $i$. $\mathrm{d} \boldsymbol{H}(\tau)=\left[\mathrm{d} H^{1}(\tau), \mathrm{d} H^{2}(\tau), \ldots, \mathrm{d} H^{M}(\tau)\right]^{\top}$ is the Radon-Nykodym differential vector of the Hawkes process, which is also a counting measure vector. The intensity of a point process can be intuitively understood as the rate of the arrivals. From (\ref{intensity of HP}), given a fixed timing $\tau$, the intensity of a Hawkes process is stimulated by an increment $\mathbf{A}\odot\Phi(t-\tau)\mathrm{d}\boldsymbol{H}(\tau)$. Now that the rate is increased due to the new arrival, the next arrival comes sooner in the probability sense. This is the so-called "self-exciting" property and the reason that the Hawkes process is commonly used to simulate the time series with clusters.

The branching matrix $\mathbf{A}$ commonly can be considered as a default setting of the whole network. With the growth of $M$, the Hawkes process can describe rather a complex network in a space sense.  However, in a temporal perspective, a constant matrix $\mathbf{A}$ is too rigid to capture the effects from the changing factors of the macro outer environment, for example, the epidemic control policies from local governments. Thus we develop an environmentally adaptive Hawkes (EAH) model with a time-varying multiplier $\mathbf{A}(t)$. 

	\section{Environmentally Adaptive Hawkes model}\label{EAH section}
In this section we develop the EAH model along with its existence, uniqueness and some interesting properties related to the p.g.fl.. Our theoretical discussion is restrained to one-dimension for convenience.
\subsection{Existence and Uniqueness of an EAH Model}
\begin{defn}[An EAH process]\label{EAH def}
	An univariate point process $\{N(t), t\geq 0\}$ is an EAH process, if it satisfies
	\begin{enumerate}
		\item $N(0) = 0$.
		\item The intensity $\lambda(t)$ is given by the Stieltjes integral
		\begin{equation}\label{intensity of EAH}
			\lambda(t) = \mu +  \alpha(t)\times \phi(t)\ast \mathrm{d}N(t),	\end{equation}
where $\mu>0$ and the environmental multiplier $\alpha(t)\geq 0$ , $\ast$ means the convolution operation and $\phi(t)\geq 0$  is the kernel function.
	\end{enumerate}
\end{defn}
In Definition \ref{EAH def}, $\alpha(t)$  serves as an characterization of the temporal evolution of the macro outer environment.  The kenel $\phi(t)$ is of the similar role as in a Hawkes process.   	Let $\alpha(t) = 1$, the EAH model reduces to a Hawkes process, which can be considered as a cluster process defined in Section 3.2 in \cite{vere1970stochastic}. The process of cluster centers    $N_{c}(t)$, is a Poisson process with rate $\mu$. The arrivals of $N_c(t)$ are often called immigrants. Each immigrant produces a subsidiary process, which is also known as a cluster. Actually, in one cluster, any individual reproduces the next generation, including the immigrant. Thus the cluster is formed by the arrivals of all-generation descendants of an immigrant. For an event arrives at $\tau$, the process of its direct descendants is constructed as a non-homogenous Poisson process with the intensity $\phi(t-\tau)$. The summation of the intensities of the process of cluster centers and all clusters gives the intensity of a Hawkes process.

Clusters in the Hawkes process are identical processes after different temporal translations. Here "identical" means that each cluster exhibits the same statistical structure leading to same properties, for example, the mean size of clusters. Define $s = \int_0^\infty \phi(\tau)\mathrm{d}\tau$, which is the mean size of a non-homogenous Poisson process with intensity $\phi(t)$, namely the mean size of the direct descendants of any event, including the immigrants.  Focus on a cluster and $s$ equals to the mean size of the first generation which is formed by the direct descendants of the immigrant. Iteratively, $s^2$ the mean size of the second generation and so on. The mean size of a cluster is the summation of all generations, namely $\sum_{n=0}^\infty s^n$, which converges with $0\leq s<1$.

By the same method as in the Hawkes process, we will construct a cluster process which possesses the intensity (\ref{intensity of EAH}) to justify the existence of EAH model. In a cluster representation of the EAH model, the "identical" property in a Hawkes process does not exist anymore. The mean size of a direct descendants process is related to the specific arrival time of the corresponding parent. Then the mean size of each generation depends on the arrivals time in the former generation iteratively. Such a high dependence on the history results in huge trouble in direct calculation of the mean cluster size. Hence we almost surely bound the mean cluster sizes to ensure the non-explosiveness, namely that the clusters are almost surely finite. Denote $$m(u)=\int_{-\infty}^{\infty} \alpha(t)\phi(t-u) d t,$$ which represents the mean size of the first generation if the ancestor of the cluster  arrives at $u$.
\begin{theorem}
	[Existence] \label{existence of cluster HT}

		Consider the intensity $\lambda(t)$ defined in (\ref{intensity of EAH}), if  there exists an $m$ ( $  0<m<1$) such that $m(u)\leq m$ for all $u>0$, then there exists a non-explosive cluster process with   $\lambda\left(t\right)$ satisfying
		$$\lambda\left(t\right)\leq \frac{\mu}{1-m} \quad a.s..$$

\end{theorem}
\begin{proof}
	See Appendix.
\end{proof}
In real applications, it is natural to have a beginning point of the process. So we clarify the uniqueness of EAH model with an arbitrarily fixed beginning point, which is straight supported by the first lemma in \cite{hawkes1974cluster}.
\begin{theorem}
	Given an arbitrarily fixed start time, there exists at most one simple point process, whose complete intensity is given by (\ref{intensity of EAH}).
\end{theorem}
\subsection{Other properties}
Now we describe the p.g.fl. of the EAH model, based on which we further derive the distribution  of the residual time of an EAH model and the length of a particular cluster in an EAH process.
 \begin{theorem}\label{pgfl of HT}
	The p.g.fl. of the EAHs has the form
	\begin{equation}
		G(z(\cdot))=\exp \left\{\int_{-\infty}^{\infty} \mu\left[F\left(z(\cdot)|t\right)-1\right] \mathrm{d} t\right\},
	\end{equation}	
where $F(z(\cdot)|t)$ is the p.g.fl. of a cluster generated by an immigrant arriving at time $t$, and including that immigrant. The functional $F(z(\cdot)|t)$ satisfies the functional equation
	\begin{equation}\label{pgfl of a cluster}
		\!\!\!\!\!\!F(z(\cdot)|t)=z(t) \exp \left\{\int_{t}^{\infty}\left[F\left(z(\cdot)|\tau\right)-1\right] \alpha(\tau)\phi(\tau-t)\mathrm{d} \tau\right\}.
	\end{equation}
\end{theorem}
\begin{proof}
	See Appendix.
\end{proof}
Now we can have an iteration formula of the forward recurrence time, also known as the residual time of an EAH model.
\begin{theorem}\label{Residual time of HT}
	The forward recurrence time, which is observed from time $y$, $L_y$ of an EAH model has survivor function
	$R_{L_y}(l)=\mathbb{P}(L_y>l)$ given by
	$$
	R_{L}(l)=\exp \left\{\mu \int_{-\infty}^{y}\left[\gamma(y, l|t)-1\right] d t-\mu l\right\},
	$$
	where $\gamma(y, l|t)$ satisfies the equation
	$$
	\begin{aligned}
			&\gamma(y, l|t)=\\&\left\{
		\begin{array}{ll}
			\exp \left\{\int_{y}^{y+l}\left[\gamma(y, l|\tau)-1\right] \alpha(\tau)\phi(\tau-t) d\tau\right\}, & y\geq t;\\
			1, & y\leq t-l;\\
			0, & \text{elsewhere}.\\
		\end{array}\right.
	\end{aligned}
	$$
\end{theorem}
\begin{proof}
	See Appendix.
\end{proof}
Also, the p.g.fl. of an EAH determines the length of a cluster, i.e. the time between the immigrant and the last individual in this cluster. If the immigrant arrives at $t$, we denote the length of the cluster as $J_t$.
\begin{theorem}\label{distribution of length of a cluster}
	The distribution function $D_{J_t}\left(y\right)=\mathbb{P}(J_t \leq y)$ of the length of a cluster, whose ancestor arrives at time $t$, satisfies, for any $y\geq 0$,
	\begin{equation}
		\begin{aligned}
					&D_{J_t}(y)=\\
					&\exp\{-m(t)
			+\int_{t}^{t+y} D_{J_\tau}(y+t-\tau)\alpha(\tau) \phi(t-\tau) \mathrm{d} \tau\}.
		\end{aligned}
	\end{equation}
\end{theorem}
\begin{proof}
	See Appendix.
\end{proof}

\subsection{Multivariate EAH and EAHDM models}
Similarly to the multivariate Hawkes process, we characterize a multivariate EAH model by its intensity defined as
\begin{equation}\label{intensity of muldim EAH}
	\boldsymbol{\lambda}(t) = \boldsymbol{\mu} +  \int_0^t \boldsymbol{\alpha}(t)\odot \Phi(t-\tau)\mathrm{d}\boldsymbol{N}(\tau), t\geq 0,
\end{equation}
 where $\boldsymbol{\alpha}(t)=(\alpha_{ij}(t))_{M\times M}$ is the environmental multipliers matrix. And the kernel function matrix, $[\Phi_{i,j}(t)]_{i,j = 1}^M$ is chosen to be an exponential kernel matrix, namely $\Phi_{i,j}(t) = \exp\{-\beta_{i,j}t\}$. For an EAH model applied on COVID-19, we reduce  our EAH to an EAHDM. Consider an $M$-dimensional model,
\begin{equation}\label{intensity of EAHDM}
	\boldsymbol{\lambda}(t) = \boldsymbol{\mu} + d(t) \int_0^t\mathbf{A}\odot \Phi(t-\tau)\mathrm{d}\boldsymbol{N}(\tau), t\geq 0,
\end{equation}
where $\mathbf{A}\in\mathbb{R}^{M\times M}$ with all entries being non-negative, and  $d(t):\mathbb{R}\to \mathbb{R}$ is a non-negative and decreasing function. In COVID-19 pandemic, the environment turned adverse to the spread of the virus due to the  alertness of the society and the  control of the government. Then this characteristic is revealed by the user-specified decaying function $d(t)$. 
\section{EM-like Algorithm}\label{EM algo}
We develop the EM-like algorithm for the EAHDM model based on the algorithm in \cite{li2017detecting}. Considering a simple point process, all arrivals are distinct   ordered by $\{t_1, t_2, \ldots, t_n\}$. The vector $\{u_1, u_2, \ldots, u_n\}$ with $u_j\in\{1,2,\ldots, M\}$ represents the corresponding dimensions (nodes) each arrival belongs to. Define auxiliary probabilities $[p_{ij}]_{n\times n}$ satisfying $\sum_{j \leq i} p_{ij} = 1 ,\forall i\in\{1,2,\ldots, n\}$.

Similarly to \cite{daley2003introduction},   let $\mathcal{L}(\mathbf{A}, d(t), \boldsymbol{\beta})$ be the log likelihood  of the multivariate EAH model.
 By the Jensen inequality in the concave case, we have 
\begin{equation}\label{lower bound of LL}
	\begin{split}
		&\mathcal{L}(\mathbf{A}, d(t), \boldsymbol{\beta})  \geq \sum_{t_i<t}p_{ii}\log\left(\frac{\mu_{u_i}}{p_{ii}}\right)\\
		&+ \sum_{t_i<t} \sum_{t_j<t_i}p_{ij}\log\left( \frac{A_{u_i,u_j}}{p_{ij}}d(t_i)\exp\{-\beta_{u_i.u_j}(t_i-t_j)\}\right)\\
		&-\sum_{i=1}^d\left(\mu_i t + \sum_{t_j<t}\frac{A_{i,u_j}d(t)}{\beta_{i,u_j}} \left(1-\exp\{-\beta_{i,u_j}(t-t_j)\}\right)\right).\\
	\end{split}
\end{equation}
Denote the right-side term of (\ref{lower bound of LL}) by  $\underline{\mathcal{L}}(\mathbf{A}, d(t), \boldsymbol{\beta};p)$, where $p$ means the set of $\{p_{ij}\}$. Given the estimation $\widehat{\mathbf{A}}^{(k)}$ from the $k$-th iteration, we can tighten the lower bound, namely $\underline{\mathcal{L}}(\mathbf{A}, \boldsymbol{\beta};p)$, using Lagrange multiplier method to get $\{\hat{p}_{ij}^{(k)}\}$.  Conditioned on $\{\hat{p}_{ij}^{(k)}\}$,   maximizing $\mathcal{L}(\mathbf{A}, d(t), \boldsymbol{\beta})$ w.r.t. $\mathbf{A}$,  then we obtain $\widehat{\mathbf{A}}^{(k+1)}$.
\begin{equation*}
	\begin{aligned}
				&\hat{p}_{ij}^{(k)} = \left\{
		\begin{aligned}
			&\frac{\mu_{u_i}}{\mu_{u_i}+\sum\limits_{j<i}\widehat{A}_{u_i,u_j}^{(k)}d(t_i)
\exp\{-\beta_{u_i,u_j}(t_i-t_j)\}}, j=i\\
			&\frac{\widehat{A}_{u_i,u_j}^{(k)}d(t_i)\exp\{-\beta_{u_i,u_j}(t_i-t_j)\}}
{\mu_{u_i}+\sum\limits_{j<i}\widehat{A}_{u_i,u_j}^{(k)}d(t_i)\exp\{-\beta_{u_i,u_j}(t_i-t_j)\} }, j<i\\
		\end{aligned}
		\right. ,\\
		&\widehat{A}_{u,v}^{(k+1)} = \frac{\beta_{u,v}\sum_{\substack{t_i<t\\u_i=u}}\sum_{\substack{t_j<t_i\\u_j=v}} \hat{p}_{ij}^{(k)}}{d(t)\sum_{\substack{t_j<t\\u_j=v}}\left(1-\exp\{-\beta_{u,v}(t-t_j)\}\right)}.\\
	\end{aligned}
\end{equation*}

To ease our implementation, we generally discretize the data by uniform binsize denoted by $\Delta$. Also we define $l = \lfloor t/\Delta \rfloor$, which is the largest integer not over $t/\Delta$ and represents the last bin. Denote the sum of arrivals in the $i$-th bin belonging to node $u$ by $n_{i,u}$. In this case, we can still introduce auxiliary probabilities and derive the corresponding EM-like algorithms. So the discretized version is
\begin{equation*}
	\begin{aligned}
		&\sum\limits_{u=1}^{M} \sum\limits_{v=1}^{M} n_{i, u} n_{j, v}\hat{p}_{ij}^{(k)}=\\&  \frac{d(\Delta i)\sum\limits_{u=1}^M\sum\limits_{v=1}^M n_{i,u} n_{j,v}\widehat{A}_{u,v}^{(k)}e^{-\Delta\beta_{u,v}(i-j)}}{\sum\limits_{u=1}^M n_{i,u}\mu_{u}+d(\Delta i)\sum\limits_{j<i}\sum\limits_{u=1}^M\sum\limits_{v=1}^M n_{i,u} n_{j,v}\widehat{A}_{u,v}^{(k)}e^{-\Delta\beta_{u,v}(i-j)}},\\
		 &\widehat{A}_{u, v}^{(k+1)}=\frac{\beta_{u,v}\sum\limits_{i\leq l}\sum\limits_{j<i}n_{i,u} n_{j,v} \hat{p}_{ij}^{(k)}}{d(\Delta l)\sum\limits_{j\leq l}n_{j,v}\left(1-\exp\{-\Delta\beta_{u,v}(l-j)\}\right)}.\\
	\end{aligned}
\end{equation*}
\section{Numerical Experiments}\label{numerical exp}

In this section, we investigate the performance of our method via numerical studies. In simulation, we will generate data from an EAHDM model and estimate $\mathbf{A}$ by the EM-like algorithm. In real data experiments, we will train a Hawkes process model and an EAHDM model with the COVID-19 pandemic data from Jan. 16 to Feb. 11 in 2020. And we will exhibit the one-step prediction with the Hawkes process model and the EAHDM model, which proves that the EAHDM model outperforms the Hawkes process. Two common cases exist where the EAHDM model will degenerate. Note that no exogenous factor steadily infects people. Hence we can assume that no immigrant process exists in the spread of COVID-19 pandemic. Only the several individuals in the beginning can be acknowledged as \textit{immigrants}, the confirmed cases afterwards were all \textit{descendants} in those clusters generated from \textit{immigrants}. So $\boldsymbol{\mu}=\mathbf{0}$ in both simulations and real data experiments.

\subsection{Simulation}
In the simulation, a $3$-dimension network is constructed. More specifically, we fix $\beta = 0.5$ and choose three $\mathbf{A}$s in which $A_{1,2}\in\{1.5, 1.8,  2.0\}, A_{2,1}\in\{1.5, 1.5, 1.5\}, A_{3,2}\in\{1.5, 1.2, 1.0\}$ and the left entries are all $0$. Based on the same method in \cite{ogata1981lewis}, we generate the data for 8 days. Since $\boldsymbol{\mu} = \mathbf{0}$, the data on the first day is manually given to trigger the whole process. Hence three nodes are equally provided $2$ arrivals on the first day, on $7$ am and $2$ pm, respectively. Then we will descritize the data with binsize $\Delta = 0.1$ to ease the further computation.  We estimate $\mathbf{A}$ by using EM-like algorithm in Section \ref{EM algo} with the known Hawkes skeleton (see \cite{embrechts2018hawkes}) imbedded. The results are shown in Table \ref{simulation results}.

\begin{table}[h!]
	\caption{Estimates of $\mathbf{A}$.}
	\begin{tabular}{ |p{4cm}p{1cm}p{1cm}p{1cm}|  }
		\hline
		& $A_{1,2}$ & $A_{2,1}$ & $A_{3,2}$\\
		\hline
		\hline
		Simulation 1 True Parameter & 1.500 & 1.500 & 1.500 \\
		Simulation 1 Estimate& 1.581  & 1.500 & 1.514 \\
		Simulation 2 True Parameter & 1.800 & 1.500 & 1.200 \\
		Simulation 2 Estimate & 1.725 & 1.500 & 1.206\\
		Simulation 3 True Parameter & 2.000 & 1.500 & 1.000\\
		Simulation 3 Estimate &1.979 & 1.500 & 0.971 \\
		\hline
	\end{tabular}
	\label{simulation results}
\end{table}

The EM-like algorithm commonly converges in  several iterations, which is rather efficient. The estimators are accurate in a mutual relationship sense but not in a absolute sense. In other words, the ratios among the estimators are robust though their values are not. Actually whether the estimates are large or small is determined by the binsize. If the binsize is chosen small, then estimates turns large. If the bins are crude, then vice versa. In Table \ref{simulation results}, we calibrate the estimates through equal $A_{2,1}$, then estimates of $A_{1,2}$ and $A_{3,2}$ are also close to the truth since the mutual relationship is estimated nicely. In real data application, we calibrate $\mathbf{A}$ by minimizing some metric between the real data and our predictions.

\subsection{COVID-19 data}
\begin{figure}[htbp]
	
	\begin{subfigure}{0.24\textwidth}
		\includegraphics[width=1\linewidth,  height=4cm]{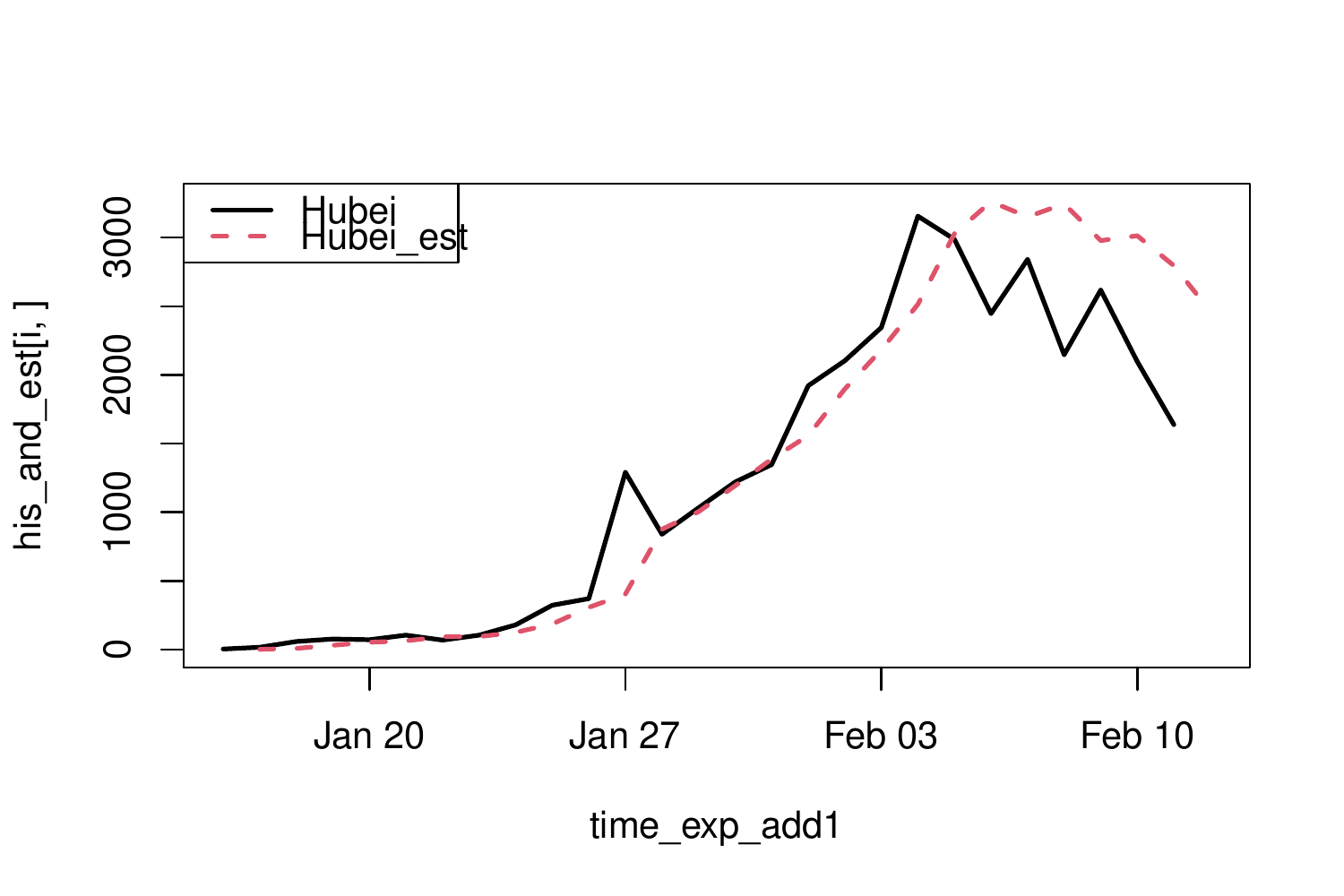}
		\caption{}
	\end{subfigure}
	\begin{subfigure}{0.24\textwidth}
		\includegraphics[width=1\linewidth,  height=4cm]{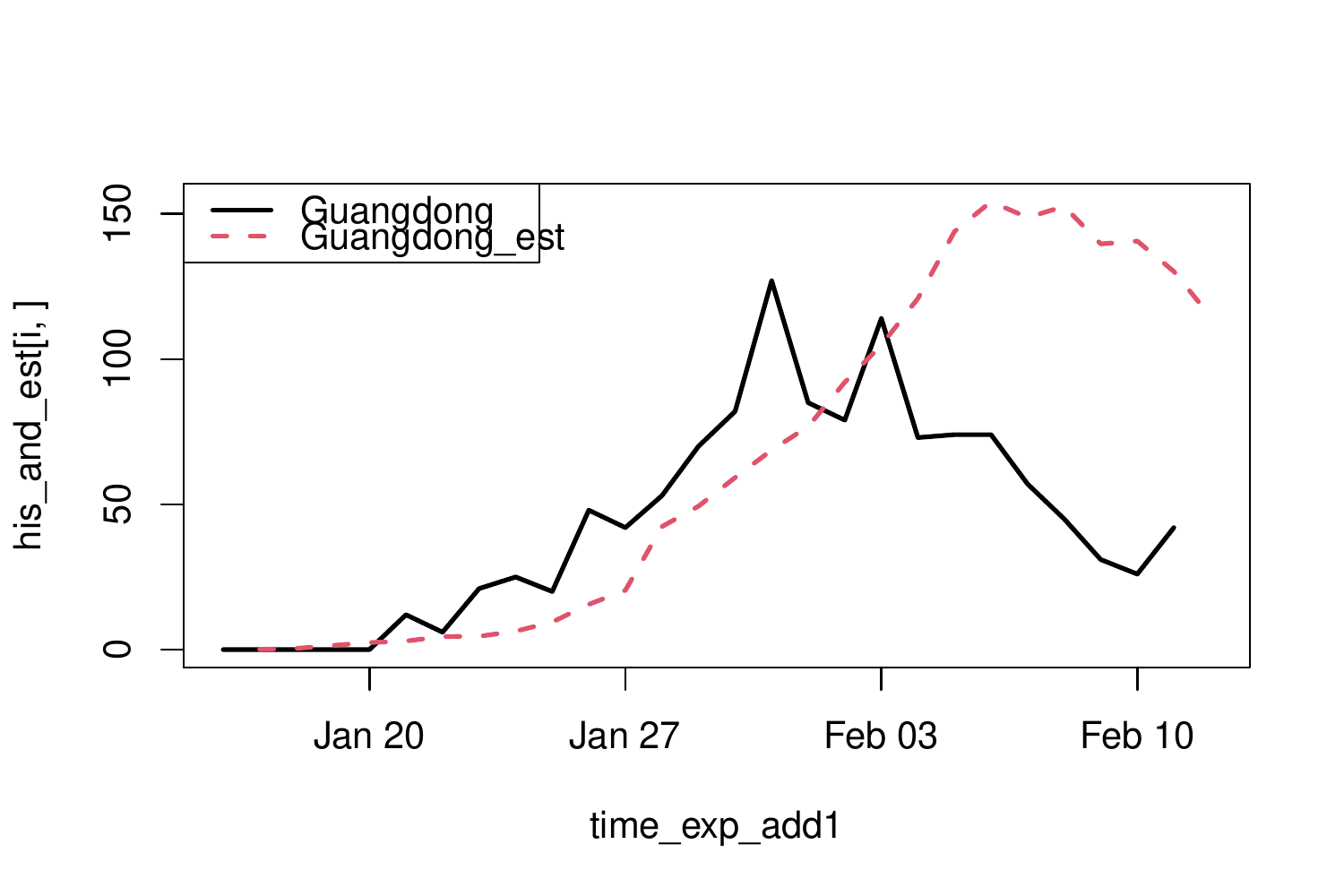}
		\caption{}
	\end{subfigure}
	
	\begin{subfigure}{0.24\textwidth}
		\includegraphics[width=1\linewidth,  height=4cm]{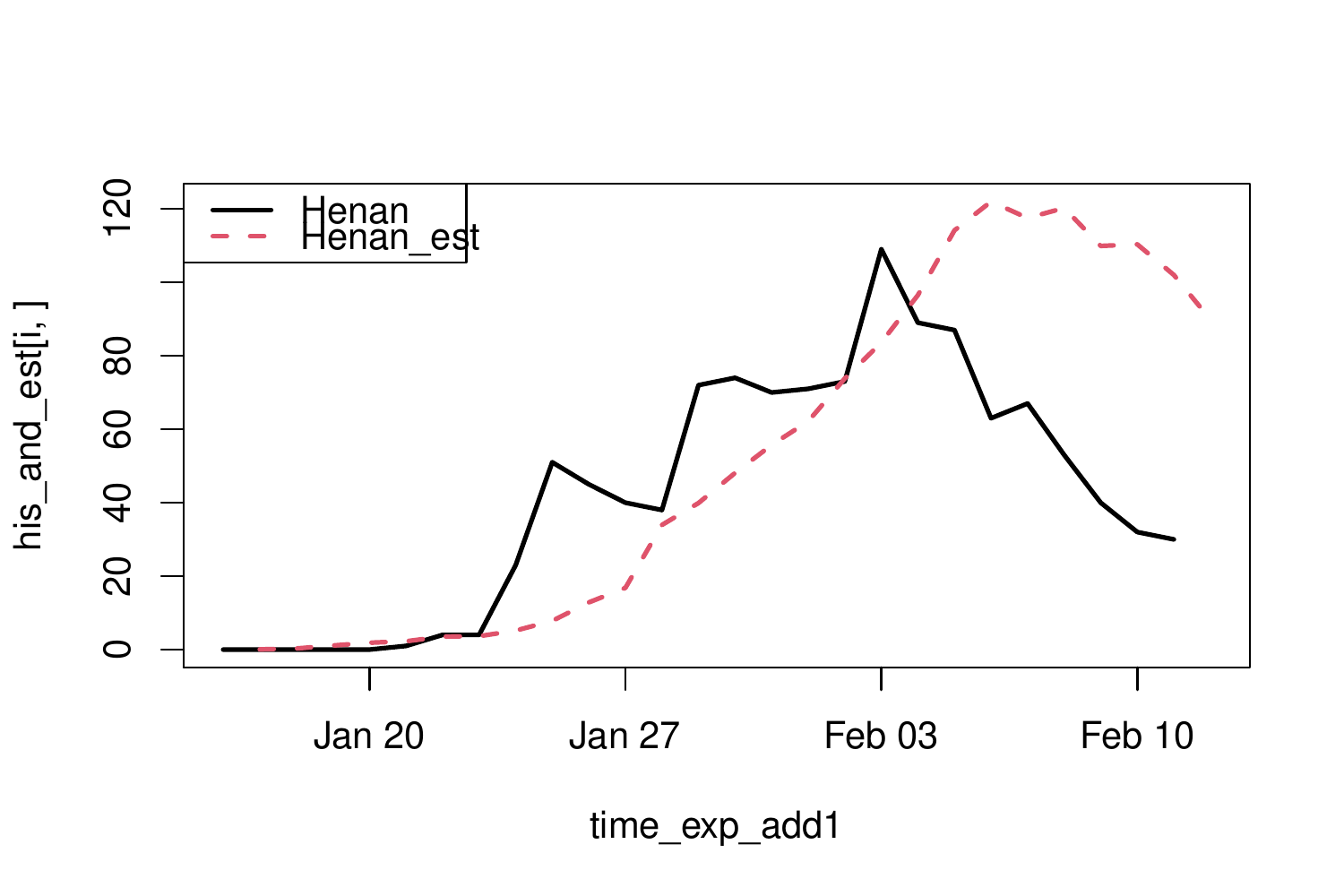}
		\caption{}
	\end{subfigure}
	\begin{subfigure}{0.24\textwidth}
		\includegraphics[width=1\linewidth,  height=4cm]{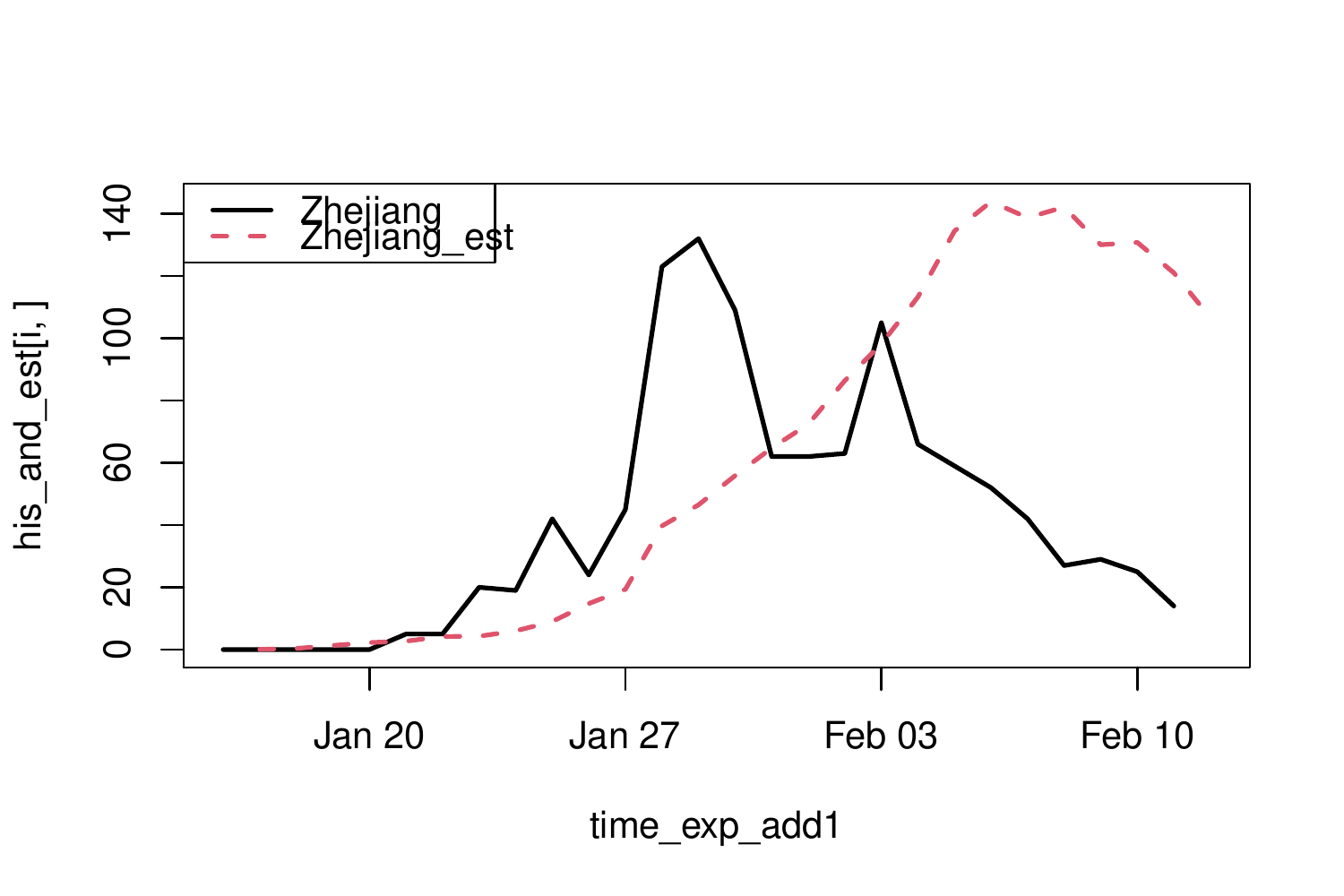}
		\caption{}
	\end{subfigure}
	
	\begin{subfigure}{0.24\textwidth}
	\includegraphics[width=1\linewidth,  height=4cm]{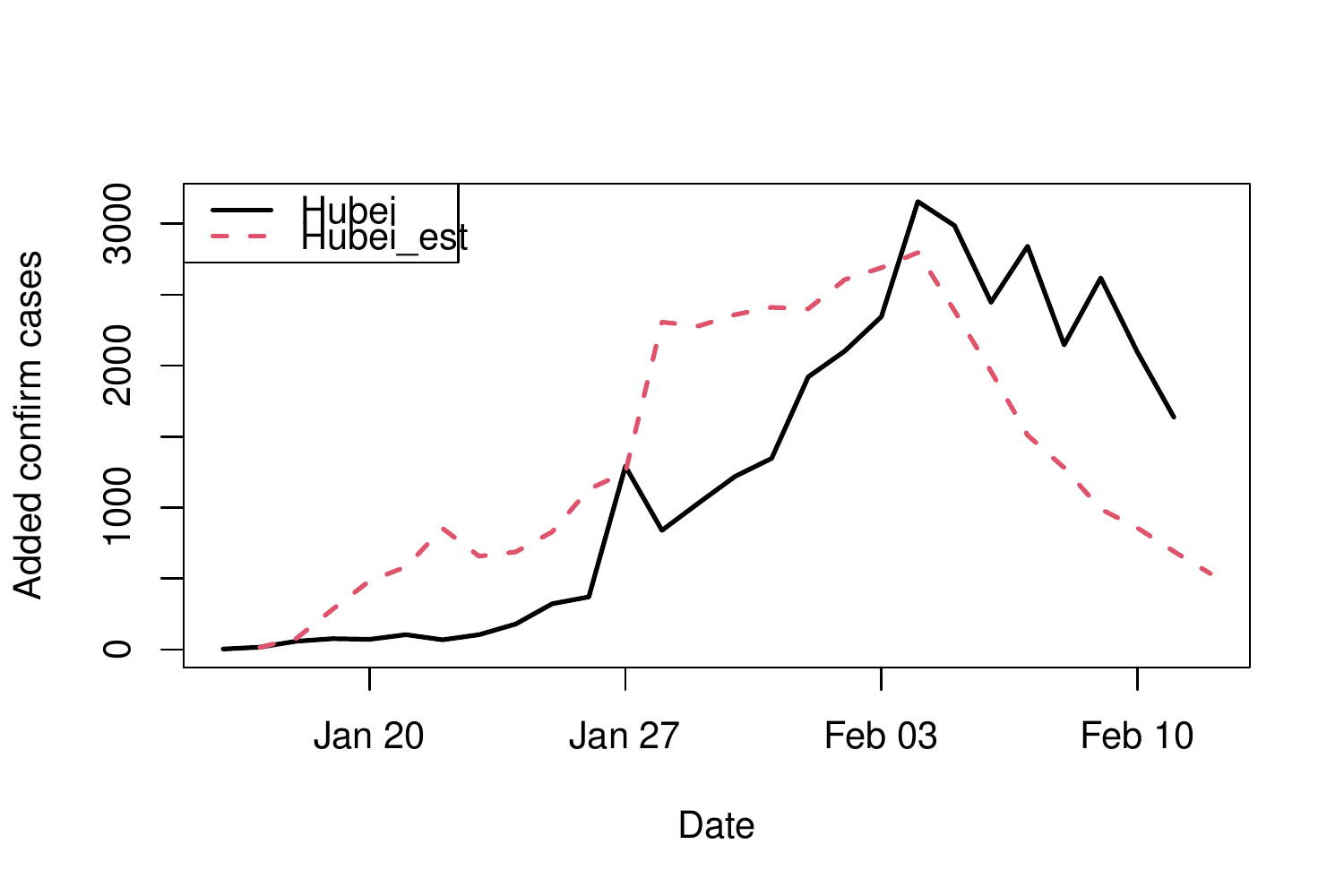}
		\caption{}
	\end{subfigure}
	\begin{subfigure}{0.24\textwidth}
	\includegraphics[width=1\linewidth,  height=4cm]{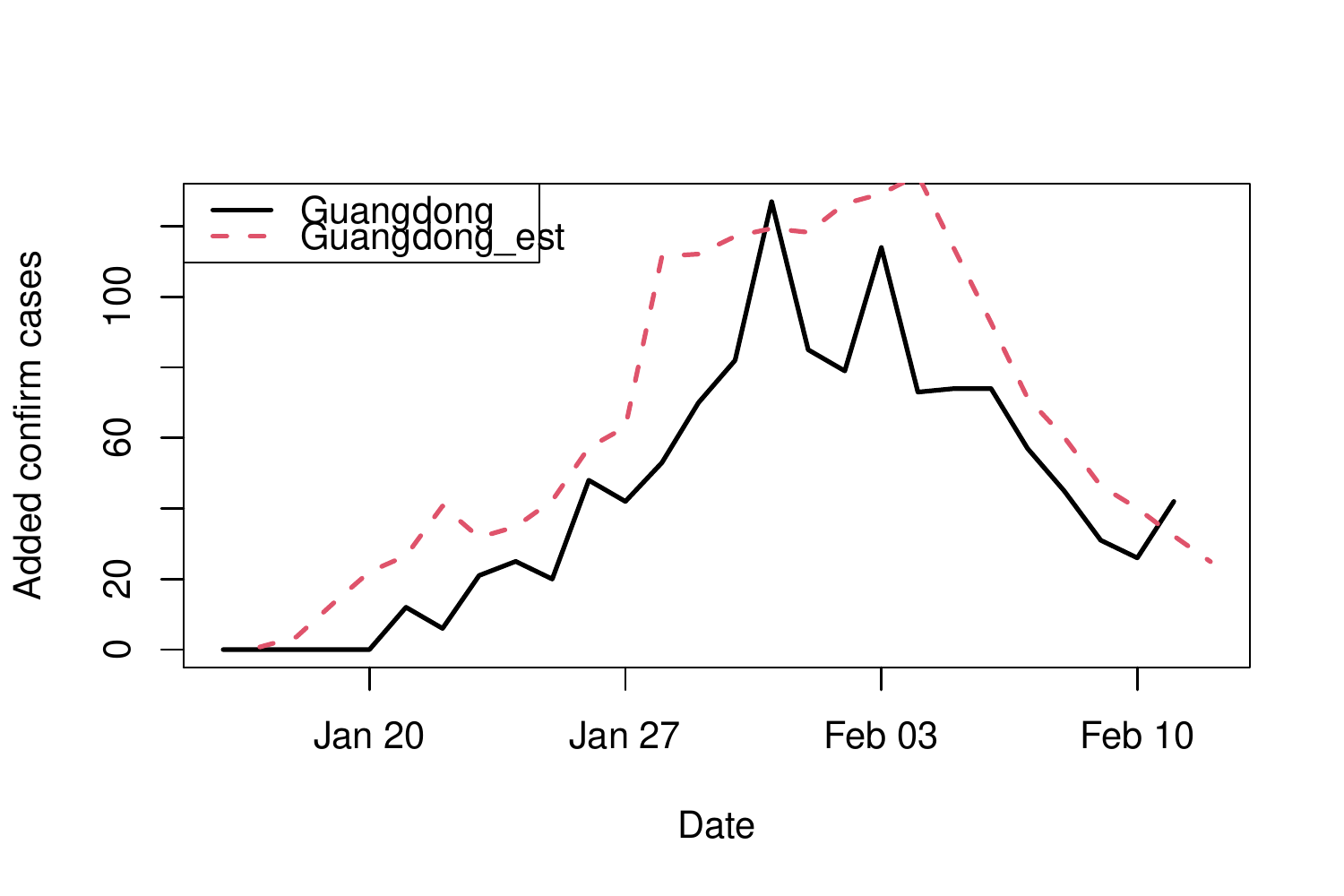}
		\caption{}
	\end{subfigure}

	\begin{subfigure}{0.24\textwidth}
		\includegraphics[width=1\linewidth,  height=4cm]{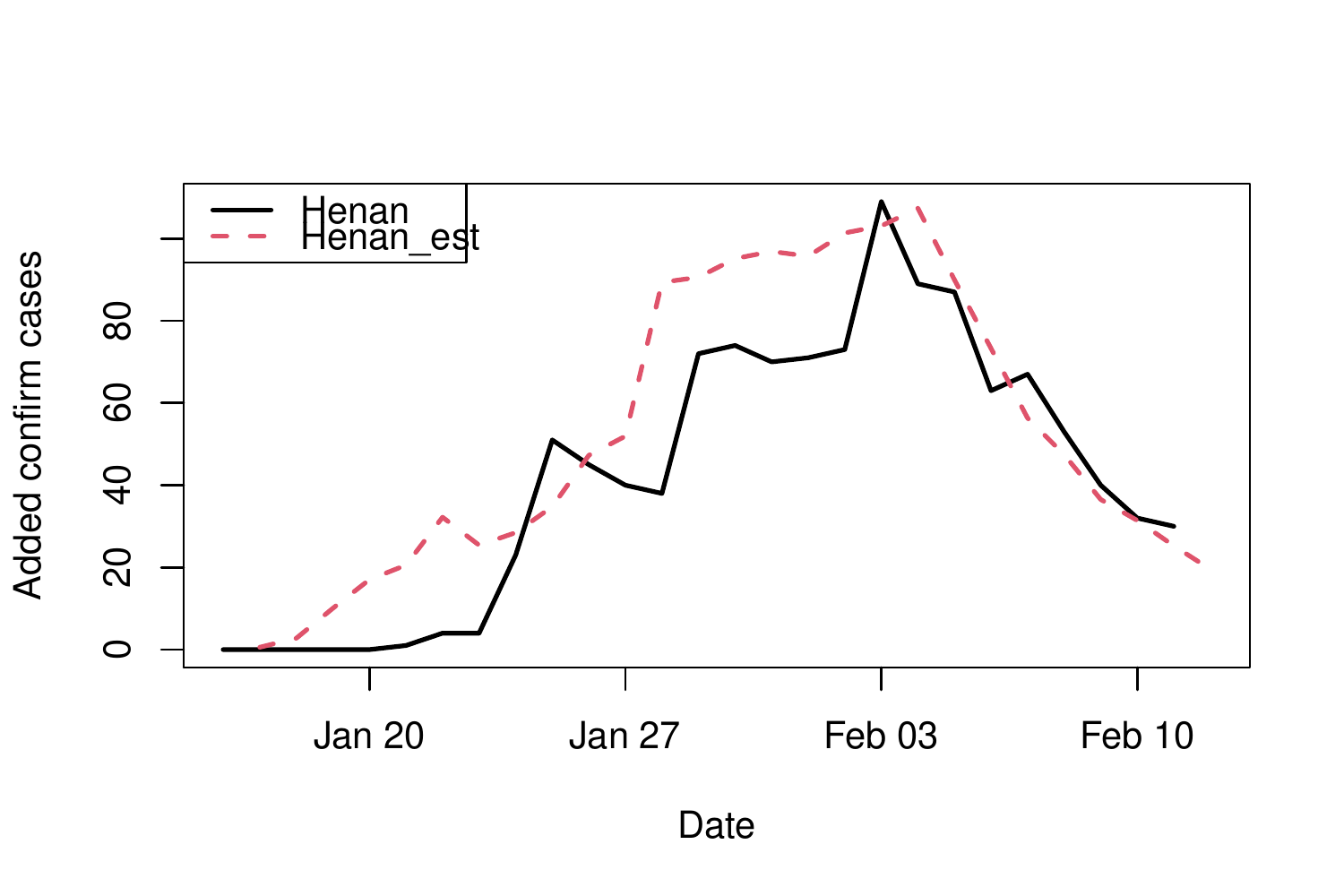}
		\caption{}
	\end{subfigure}
	\begin{subfigure}{0.24\textwidth}
		\includegraphics[width=1\linewidth,  height=4cm]{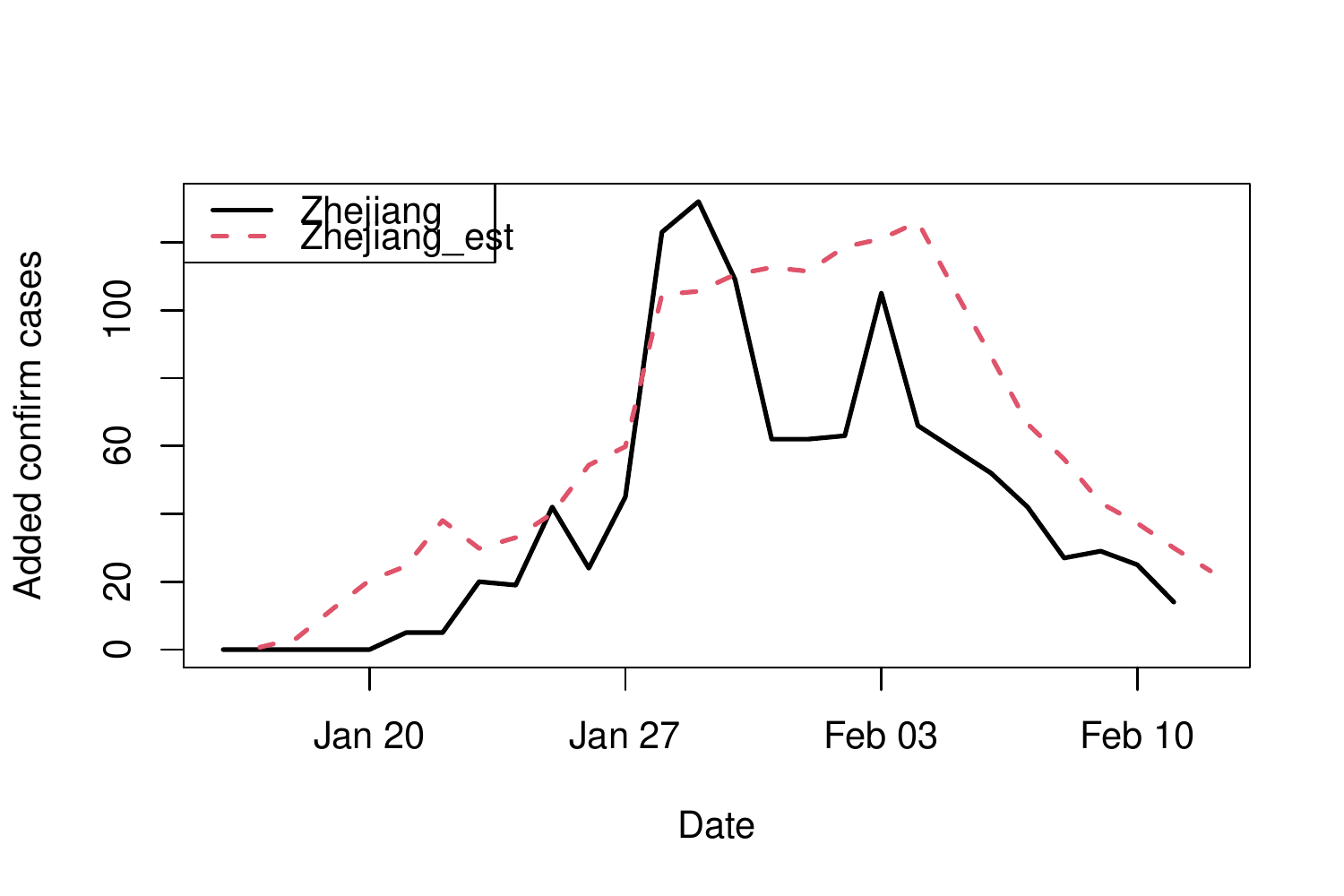}
		\caption{}
	\end{subfigure}

\caption{One-step estimate on the network formed by Hubei, Guangdong, Zhejiang and Henan. $\beta = 0.5$. The black lines represent the real data and the red dotted lines represent the one-step predictions. Subfigures (a-d) are from the Hawkes process. Subfigures (e-h) are from the EAHDM.}
\label{beta 0.5}
\end{figure}

\begin{figure}[htbp]
	
	\begin{subfigure}{0.24\textwidth}
		\includegraphics[width=1\linewidth,  height=4cm]{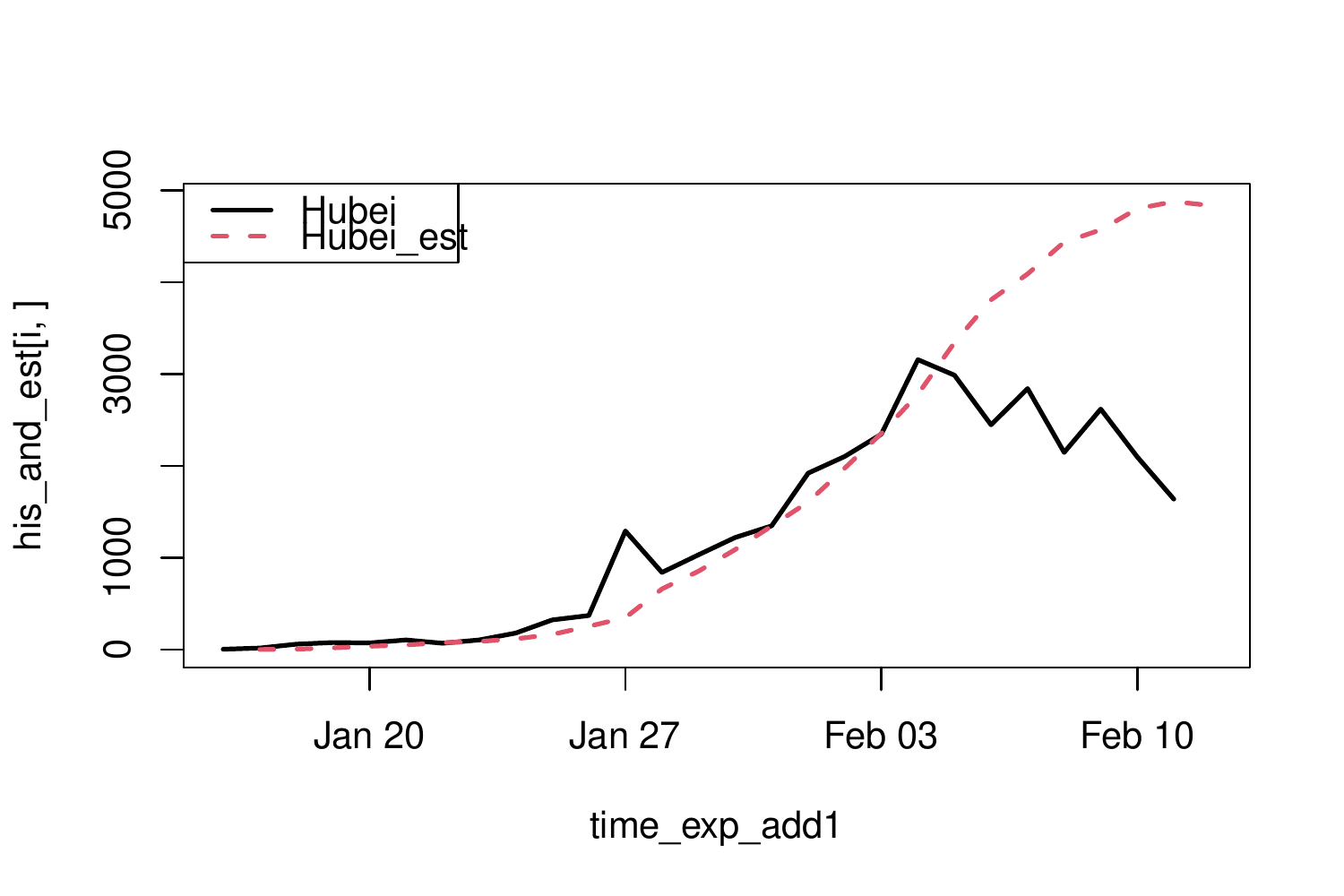}
		\caption{}
	\end{subfigure}
	\begin{subfigure}{0.24\textwidth}
		\includegraphics[width=1\linewidth,  height=4cm]{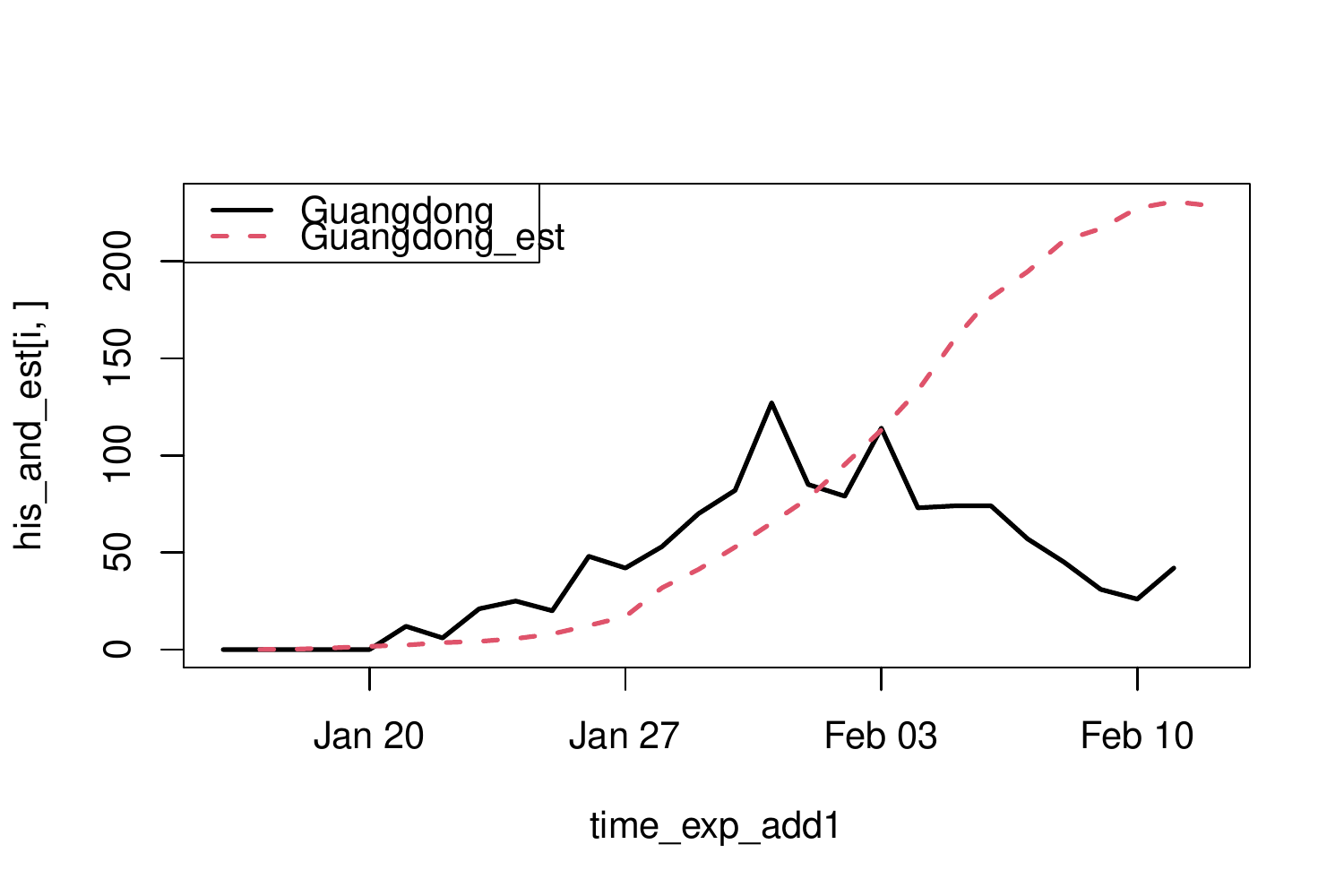}
		\caption{}
	\end{subfigure}
	
	\begin{subfigure}{0.24\textwidth}
		\includegraphics[width=1\linewidth,  height=4cm]{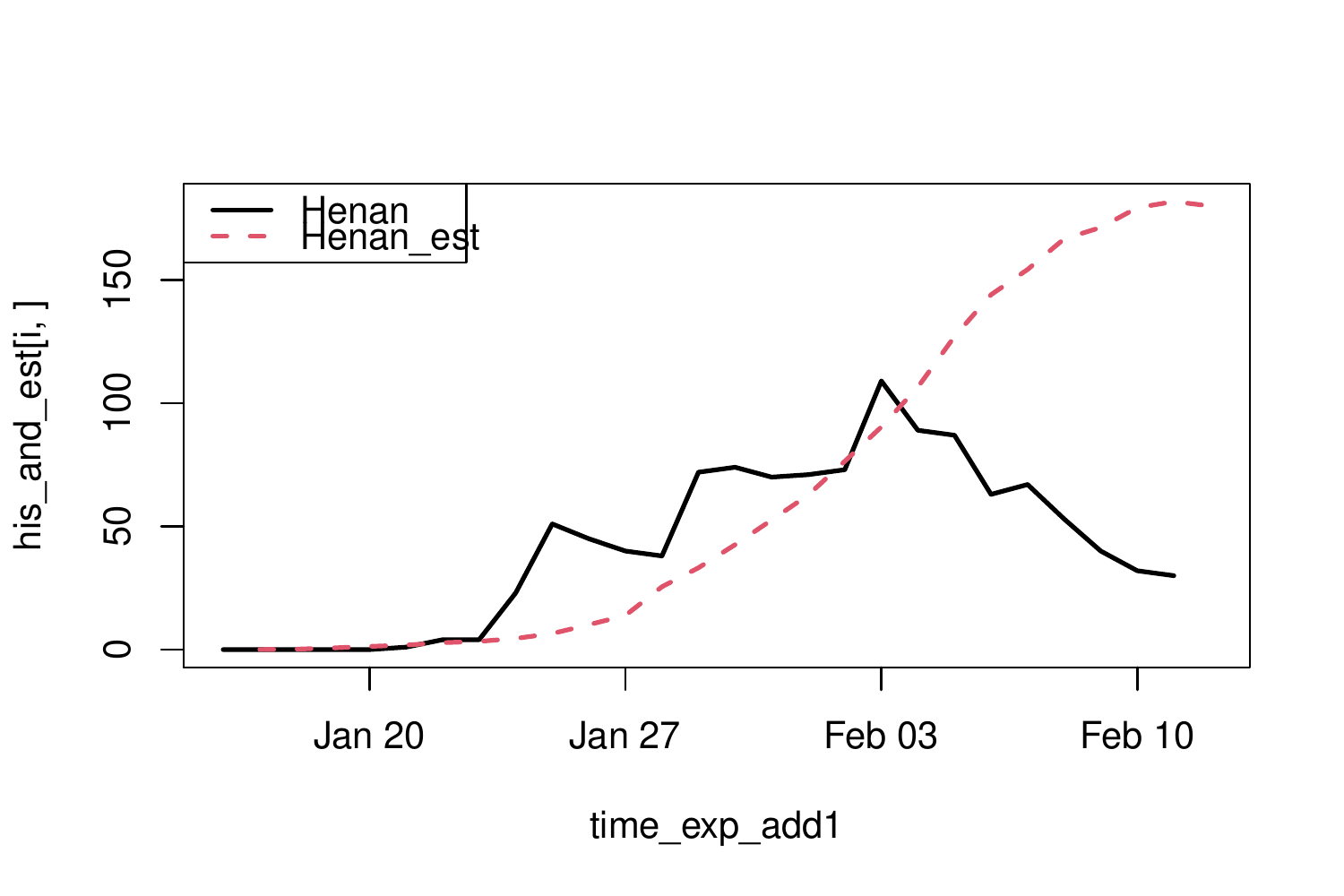}
		\caption{}
	\end{subfigure}
	\begin{subfigure}{0.24\textwidth}
		\includegraphics[width=1\linewidth,  height=4cm]{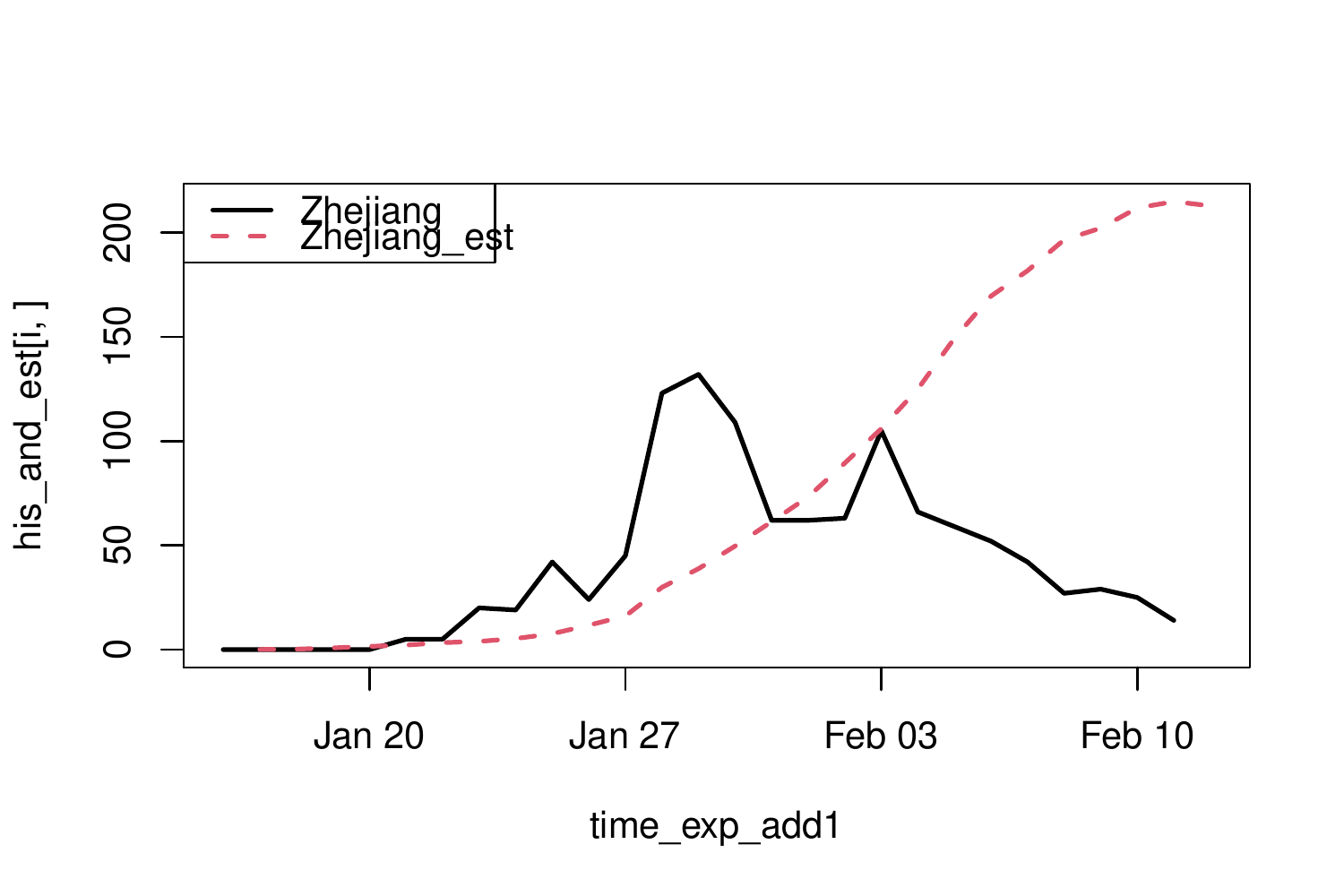}
		\caption{}
	\end{subfigure}
	
	\begin{subfigure}{0.24\textwidth}
		\includegraphics[width=1\linewidth,  height=4cm]{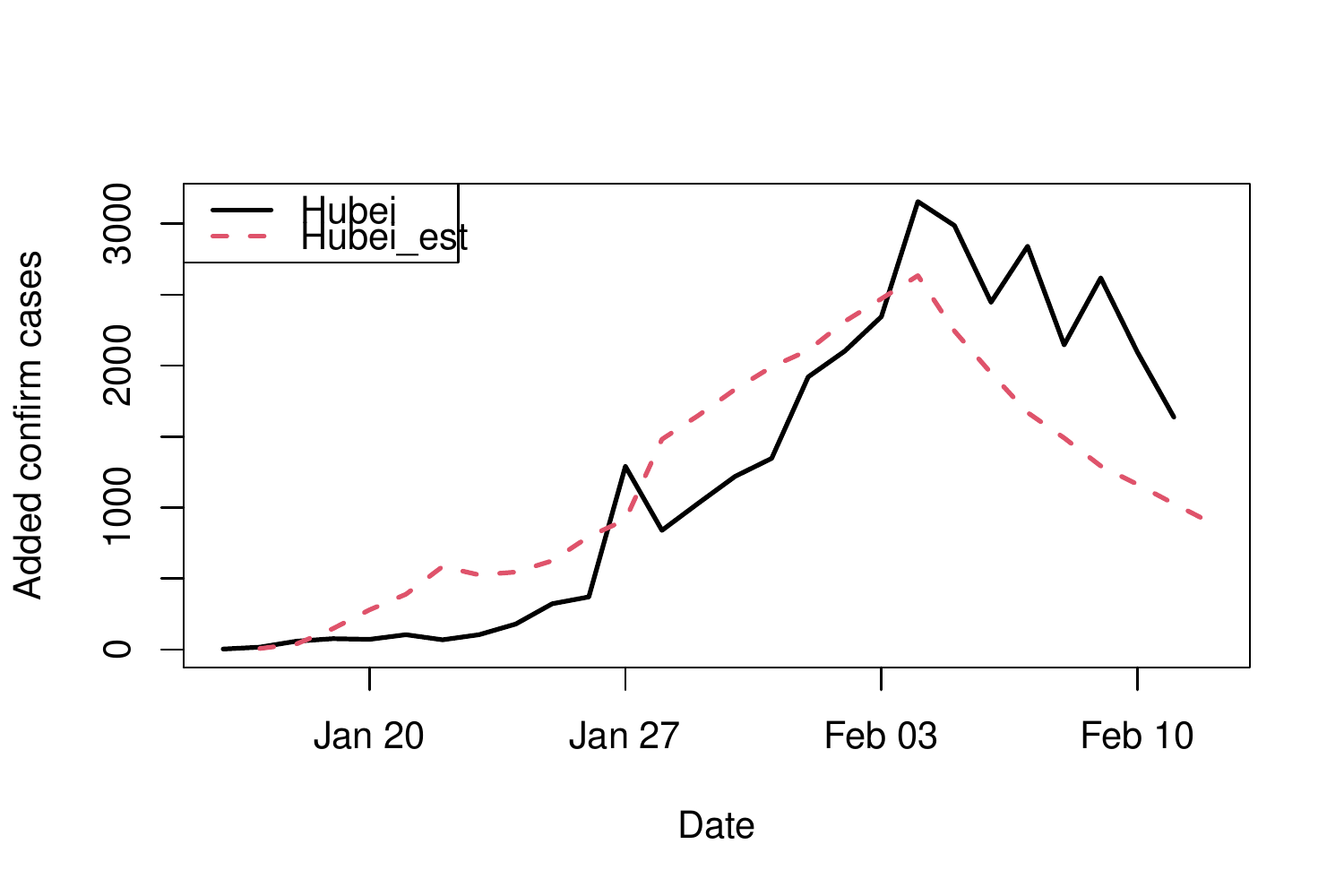}
		\caption{}
	\end{subfigure}
	\begin{subfigure}{0.24\textwidth}
		\includegraphics[width=1\linewidth,  height=4cm]{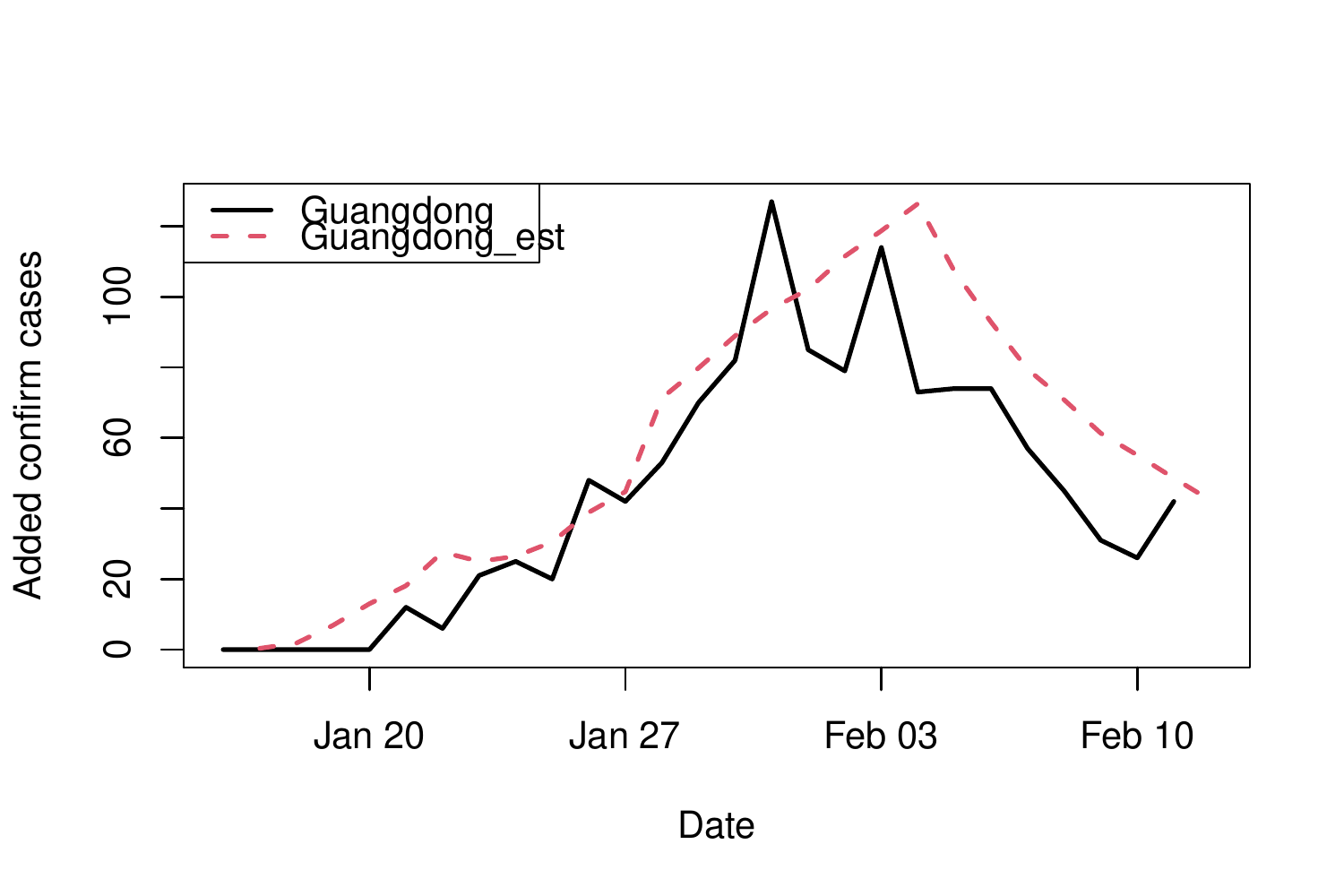}
		\caption{}
	\end{subfigure}
	
	\begin{subfigure}{0.24\textwidth}
		\includegraphics[width=1\linewidth,  height=4cm]{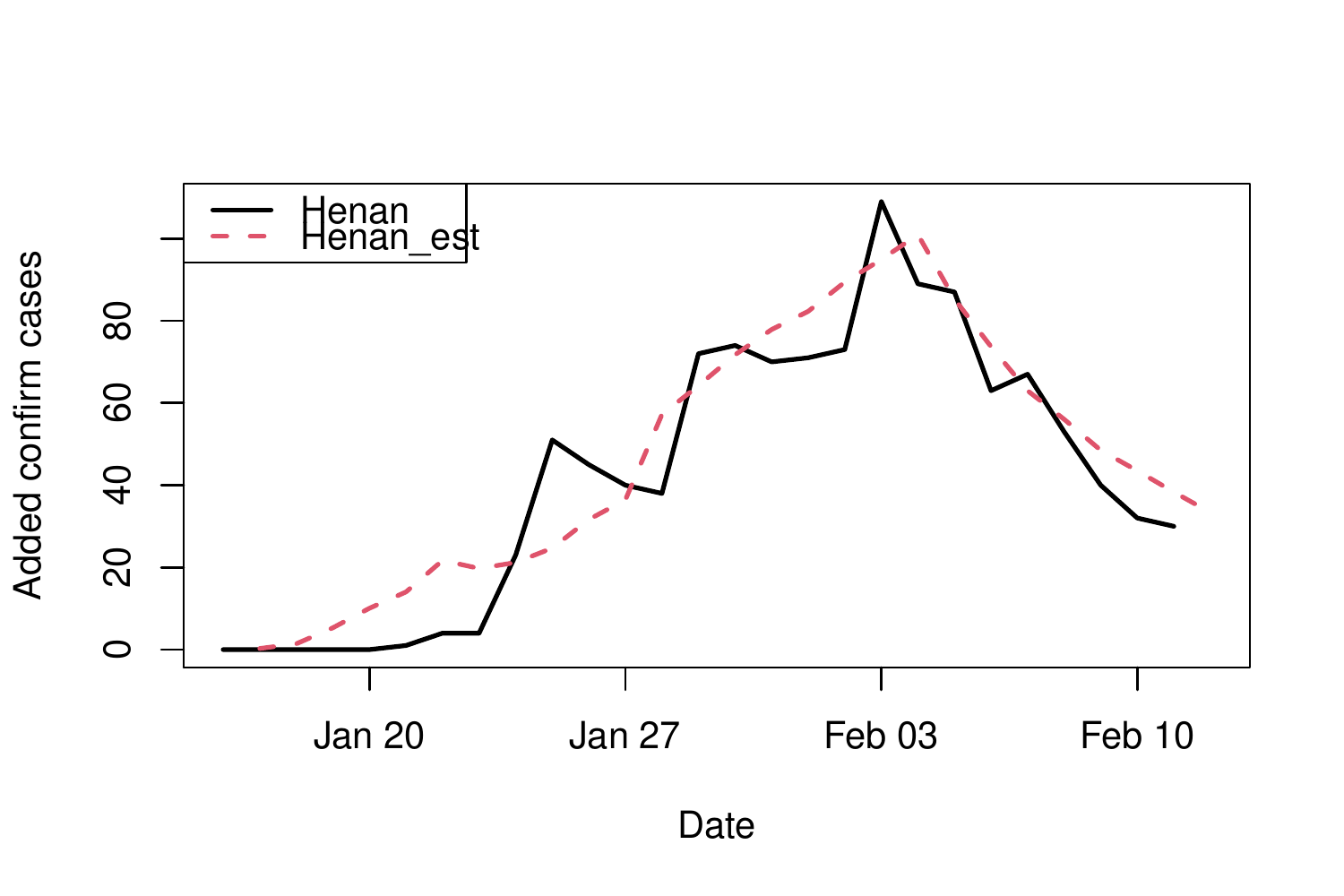}
		\caption{}
	\end{subfigure}
	\begin{subfigure}{0.24\textwidth}
		\includegraphics[width=1\linewidth,  height=4cm]{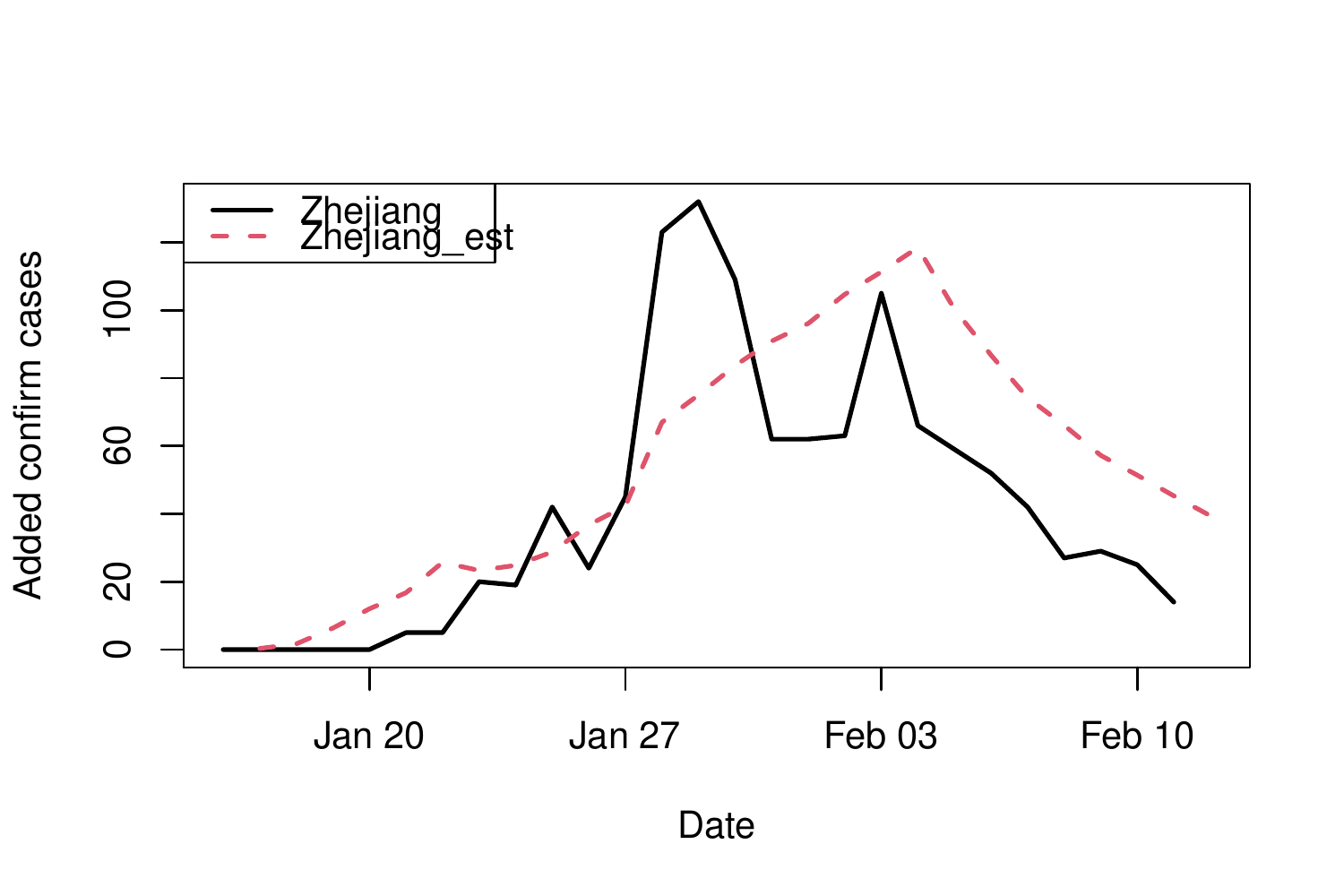}
		\caption{}
	\end{subfigure}
	
\caption{One-step estimate on the network formed by Hubei, Guangdong, Zhejiang and Henan. $\beta = 0.1$. The black lines represent the real data and the red dotted lines represent the one-step predictions. Subfigures (a-d) are from the Hawkes process. Subfigures (e-h) are from the EAHDM.}
\label{beta 0.1}
\end{figure}

We now utilize the Hawkes processes and the EAHDM model to fit the COVID-19 data in China and then make one-step predictions. Our data is obtained from \cite{wu2020open} and we choose the data source National Health Commission of Chinese. The period is from Jan. 16 to Feb. 11 in 2020  containing added confirm cases in four provinces. Those four places are Hubei, Guangdong, Zhejiang and Henan, whose pandemic were once the most severe.  In many implementation work involved the Hawkes process, $\beta$ is set manually (for example in  \cite{li2017detecting}). In our experiments, we let $\beta\in\{0.5, 0.1\}$. In the EAHDM model,
$$
d(t) = \left\{\begin{aligned}
	&1/\max(7,t)^2, \quad 0\leq t\leq 20\\
	&1/(t^{2.4}-926.7), \quad t>20\\
\end{aligned}\right..
$$
The results shown in   Figures \ref{beta 0.5} and  \ref{beta 0.1} indicate that the static branching matrix $\mathbf{A}$ in the Hawkes process makes it hard to capture the evolution of the environment temporally. Thus the one-step predictions in the Hawkes process overestimate the severity after about Feb. 3. In the EAHDM model, the drawback of the Hawkes process is obviously fixed. Also, as $\beta$ becomes smaller (from $0.5$ to $0.1$), the overestimates in the Hawkes process get worse, while the EAHDM model behaves more robust. It is interesting   that if $\beta = 1$, the Hawkes process will ease its overestimates, but still not as good as the EAHDM model. Additionally, it is not recommended to choose a $\beta$ even larger than $0.5$. Compared to $d(t)$ which reflects the evolution of outer environment, $\beta$ means the decay rate of the influence from history events which can be interpreted as the infectivity of COVID-19 virus. If $\beta = 1$, then $\exp\{-\beta t\}$ will be too small to capture the influence of the history only after $5$ days, which is not reasonable according to our common perception to the COVID-19 virus.

\section{Conclusion}
The Hawkes process is a classical self-exciting process which is frequently used to model the data with clusters. Considering the Hawkes process lacks of the flexibility in a environmentally temporal sense,  we develop a novel and sophisticated EAH model and clarify its existence, uniqueness and the non-explosive condition. In implementation stage, we reduced the EAH model to the EAHDM model, which is still an extension of the Hawkes process. In simulations, the EM-like algorithm requires a few iterations and provides nice estimates of the mutual relationship between nodes in a network. In the real data experiments, a proper decay function $d(t)$ displays that the EAHDM outperforms the Hawkes process when predicting the evolution of the COVID-19 pandemic in China. Also, the EAHDM model is more robust with the manually chosen parameter $\beta$  than  the Hawkes process.

\section*{Acknowledgment}

We are indebted to the National Science Foundation
of China (Nos. 71771203, 11671374). We are  grateful to Prof. Yao Xie for helpful discussion.

\bibliographystyle{IEEEtran}
\bibliography{refs}

\appendix

\textbf{Proof of Theorem \ref{existence of cluster HT}:}
\begin{proof}
	
	The first step is to construct a cluster process whose intensity is exactly the form in (\ref{intensity of EAH}). The process of cluster centers $N_{c}(t)$ is set to be a Poisson process of rate $\mu$. For each individual arriving at time $\tau$, the process of the arrivals of its direct descendants is constructed as a non-homogenous Poisson process with the intensity $\alpha(t)\phi(t-\tau)$.
	
	So far we construct the cluster process we want intuitively. If the existence of such a cluster process is ensured, based on the fact that the cluster process is the superposition of every clusters and the process of cluster centers, it is straightforward to see the constructed process possesses exactly the intensity in (\ref{intensity of EAH}).
	
	So in the second step we try to ensure the existence of constructed process and control its rate. Now that the process of cluster centers is a homogenous Poisson process, which is relatively easy to study with, we focus on the size of each cluster. From Chapter 1 in \cite{harris1964theory},  particularly Theorem 5.1, we may firstly bound the size of first generation and the size of other generations can be controled by a simple iteration.
	
	WLOG we consider a immigrant generated by $N_c(\cdot)$ at time $u$. As said above, the first generation directly generated by the immigrant follows a non-stationary $\alpha(t)\phi(t-u),t>u$ Poisson process. In Section 4.1 of \cite{vere1970stochastic}, we recall the p.g.fl of non-homogeneous Poisson process as
	\begin{equation}
			G(\xi(\cdot)|u)
			=\exp \left\{\int_{}^{}(\xi(t)-1) \alpha(t)\phi(t-u) d t\right\}.
	\end{equation}
	
	Now we truncate the cluster process at the first generation (including the first generation). Since p.g.fl. provides sufficient characterization of a point process,
	Section 4.2 of \cite{vere1970stochastic} indicates that the process of cluster centers with its first generation descendents is equivalent to a Neyman-Scott cluster process. Actually,
	$$
	G(\xi(\cdot)|u)
	=\exp \left\{m(u)\int_{}^{}\frac{ \alpha(t)\phi(t-u)}{m(u)}(\xi(t)-1) d t\right\}.
	$$
	The cluster size in such a Neyman-Scott process, namely the size of first generation of a cluster, whose ancestor arrived at time $u$, in the formerly constructed process, has a Poisson distribution with mean $m(u)$. Namely, the size of the first generation of a cluster whose ancestor arriving at time $u$ obeys $\operatorname{Poi}(m(u))$.

	Now the cluster in the constructed process contains all generations, by Theorem 5.1 in \cite{harris1964theory}, we can calculate the upper bound of mean cluster size with probability one as,
	
	\begin{equation*}
		\sum_{n=0}^\infty m^n = \frac{1}{1-m}.
	\end{equation*}
	
	Meanwhile, by Theorem 6.1 in \cite{harris1964theory}, when $m<1$, the cluster is a.s. finite. As a result, Corollary 3.2 along with Theorem 3 in \cite{westcott1971existence}  ensures the existence of this non-explosive cluster process. Now ends the proof.
\end{proof}

\textbf{Proof of Theorem \ref{pgfl of HT}}

\begin{proof}
	By the first formula in Section 3.2 of \cite{vere1970stochastic}, we have
	\begin{equation*}
		G(z(\cdot)) = G_0\left(F(z(\cdot)|t)\right),
	\end{equation*}
	where $G_0(z(\cdot))$ is the p.g.fl. of a homogenous Poisson process with intensity $\mu$. By the first formula in Section 4.1 of \cite{vere1970stochastic}, $G_0(z(\cdot))$ is rather clear to be
	\begin{equation*}
		G_0(z(\cdot))=\exp \left\{\int_{-\infty}^\infty\mu(z(t)-1) \mathrm{d}t\right\}.
	\end{equation*}
	
	We only need to figure out $F(z(\cdot)|t)$. We define $F_n(z(\cdot)|t)$ to be the p.g.fl of the cluster which consists of all births in all generations up to and including the $n$-th generation descendants with the immigrant arriving at $t$.
	
	Denote $H(z(\cdot)|t)$ is the p.g.fl. of a non-homogenous Poisson process, by which an individual arrives at time $t$ generates its descendants, in a EAH model. So actually
	\begin{equation*}
		H(z(\cdot)|t) = \exp\left\{\int_t^\infty (z(\tau)-1)\alpha(\tau)\phi(t-\tau)\mathrm{d}\tau\right\}.
	\end{equation*}
	
	Then again, similarly with $G(z(\cdot)) = G_0\left(F(z(\cdot)|t)\right)$, by the smoothing formula of p.g.fl.,
	\begin{equation*}
		\begin{split}
			&F_n(z(\cdot)|t)\\ &= z(t) H\left(F_{n-1}(z(\cdot)|\tau)\right|t)\\
			&=z(t)\exp\left\{\int_t^\infty \left(F_{n-1}(z(\cdot)|\tau)-1\right)\alpha(\tau)\phi(\tau-t)\mathrm{d}\tau\right\}.\\
		\end{split}
	\end{equation*}
	Here $F_0(z(\cdot)|t) = \mathbb{E}(z(t)|t) = z(t)$ represents the ancestor of the cluster, which is also the immigrant arriving at time $t$. Let $n\to\infty$ and the proof ends.
\end{proof}

\textbf{Proof of Theorem \ref{Residual time of HT}}

\begin{proof}
	If we take
	\begin{equation}
		z(x)=\left\{\begin{array}{ll}
			z & y\leq x\leq y+l \\
			1 & elsewhere\\
		\end{array}\right.,
	\end{equation}
	then
	\begin{equation}
		\begin{aligned}
			F\left(z(\cdot)|t\right)
			&=\mathbb{E}\left(e^{\int\log(z(u))dN_s(u|t)}\right)\\
			&=\mathbb{E}\left(e^{\int_{y}^{y+l}zdN_s(u|t)}\right)\\
			&=\mathbb{E}\left(z^{N_s\left((y,y+l)|t\right)}\right)\\
			&=:\pi(y, l, z|t),
		\end{aligned}
	\end{equation}
	say is the p.g.f. for the number of events in the interval $(y, y+l)$ for a cluster whose originating event occurs at time $t$.
	It is easy to see
	$$
	\pi(y,l,z|t)=\left\{\begin{array}{ll}
		\mathbb{E}\left(z^{N_s\left((y,y+l)|t\right)}\right) & t\leq y\\
		\mathbb{E}\left(z^{N_s\left((t,y+l)|t\right)}\right) & y<t<y+l \\
		1 & t\geq y+l\\
	\end{array}\right..
	$$
	Again, based on (\ref{pgfl of a cluster}),
	\begin{equation}
		\begin{aligned}
			&\pi(y,l,z|t)
			=\\&\left\{\begin{array}{ll}
				\exp \left\{\int_{y}^{y+l}\left(\pi\left(y,l,z|\tau\right)-1\right) \alpha(\tau)\phi(\tau-t)\mathrm{d} \tau\right\} & t\leq y \\
				1 & t\geq y+l\\
				z \exp \left\{\int_{t}^{y+l}\left(\pi\left(y,l,z|\tau\right)-1\right) \alpha(\tau)\phi(\tau-t)\mathrm{d} \tau\right\} & \text{elsewhere}
			\end{array}\right.
		\end{aligned}
	\end{equation}
	Then
	\begin{equation}
		\begin{aligned}
			G(z(\cdot))&= \exp \left\{\int_{-\infty}^{\infty} \mu\left(F(z(\cdot)|t)-1\right) d t\right\}\\
			&= \exp \left\{\int_{-\infty}^{\infty} \mu\left(\pi(y,l,z|t)-1\right) d t\right\}\\
			&= \exp \left\{\int_{-\infty}^{y+l} \mu\left(\pi(y,l,z|t)-1\right) d t\right\}\\
			&=:Q(y,l,z).\\
		\end{aligned}
	\end{equation}
	Now we look at the distribution of the residual time of a EAH,
	$$
	\begin{aligned}
		&\mathbb{P}(L_y>l)\\ &= \mathbb{P}(\text{Start observing from } y, \text{no events happen in }[t, t+l))\\
		&= \mathbb{E}(0^{N(y,y+l)})\\
		& = Q(y,l,0) \\
		& = \exp \left\{\int_{-\infty}^{y+l} \mu\left[\pi(y,l,0|t)-1\right] d t\right\}.\\
	\end{aligned}
	$$
	Since
	\begin{equation}
		\begin{aligned}
			&\pi(y,l,0|t)
			=\\&\left\{\begin{array}{ll}
				\exp \left\{\int_{y}^{y+l}\left(\pi\left(y,l,0|\tau\right)-1\right) \alpha(\tau)\phi(\tau-t)\mathrm{d} \tau\right\} & t\leq y\\
				1 & t\geq y+l\\
				0 & \text{elsewhere} \\
			\end{array}\right..
		\end{aligned}
	\end{equation}
	Then
	$$
	\begin{aligned}
		\mathbb{P}(L_y>l)
		& = \exp \left\{\int_{-\infty}^{y+l} \mu\left[\pi(y,l,0|t)-1\right] d t\right\}\\
		&=\exp \left\{\int_{-\infty}^{y} \mu\left[\pi(y,l,0|t)-1\right] d t-\mu l\right\}.\\
	\end{aligned}
	$$
	We define $\gamma(y,l|t) = \pi(y,l,0|t)$ and the proof ends.
\end{proof}

\textbf{Proof of Theorem \ref{distribution of length of a cluster}}

\begin{proof}
	Let
	$$
	z(x)=\left\{
	\begin{array}{ll}
		1 & x\leq y\\
		0 & x>y\\
	\end{array}\right..
	$$
	Then
	$$
	\begin{aligned}
		F(z(\cdot)|t)=&\mathbb{E}\left(\exp \left\{ \int_{-\infty}^{\infty} \log z(\tau) d N_s(\tau|t)\right\}\right) \\=& \mathbb{E}(0^{N_s((y,\infty)|t)}) = \mathbb{P}(N_s((y,\infty)|t) = 0).
	\end{aligned}
	$$
	Also we find that
	$$
	D_{J_t}(y) = \mathbb{P}(J_t\leq y) = \mathbb{P}(N_s((t+y,\infty)|t)=0).
	$$
	Based on (\ref{pgfl of a cluster}), for $y\geq 0$,
	$$
	\begin{aligned}
		D_{J_t}(y)
		&=\exp\left\{\int_{t}^{\infty}\left(D_{J_\tau}(y+t-\tau)-1\right) \alpha(\tau)\phi(\tau-t) \mathrm{d} \tau\right\}\\
		&=\exp \left\{-m(t)+\int_{t}^{t+y} D_{J_\tau}(y+t-\tau)\alpha(\tau) \phi(\tau-t) \mathrm{d} \tau\right\}.\\
	\end{aligned}
	$$
\end{proof}

\end{document}